\documentclass{iau}
\usepackage{graphicx}
\DeclareGraphicsExtensions{.pdf,.png,.jpg,.eps}
\usepackage{caption}
\usepackage{wrapfig}
\def\be{\begin{equation}}
\def\ee{\end{equation}}
\title[VA14.~~Low/high redshift galaxy separation] 
{Low/High Redshift Classification of Emission Line
Galaxies in the HETDEX survey}

\author[Acquaviva$^1$, Gawiser$^2$, Leung$^2$ \& Martin$^1$]   
{Viviana Acquaviva$^1$, Eric Gawiser$^2$, Andrew S. Leung$^2$ \\ and Mario R. Martin$^1$}
\affiliation{$^1$ Physics Department, CUNY New York City College of Technology, Brooklyn, NY 11201 \\ email: {\tt vacquaviva@citytech.cuny.edu} \\[\affilskip]
$^2$ Department of Physics and Astronomy, Rutgers, The State University of New Jersey, Piscataway, NJ 08554}

\pubyear{2014}
\volume{306}  
\jname{Statistical Challenges in 21st Century Cosmology}
\editors{Alan Heavens, Jean-Luc Starck, Alberto Krone-Martins, eds.}
\begin{document}

\maketitle

\begin{abstract}
We discuss different methods to separate high- from low-redshift galaxies
based on a combination of spectroscopic and photometric observations. Our
baseline scenario is the Hobby-Eberly Telescope Dark Energy eXperiment
(HETDEX) survey, which will observe several hundred thousand Lyman Alpha
Emitting (LAE) galaxies at 1.9 $<$ z $<$ 3.5, and for which the main source of
contamination is [OII]-emitting galaxies at z $<$ 0.5. Additional information useful for
the separation comes from empirical knowledge of LAE and [OII] luminosity
functions and equivalent width distributions as a function of redshift. We consider
three separation techniques: a simple cut in equivalent width,
a Bayesian separation method, and machine learning algorithms, including
support vector machines. These methods can be easily applied to
other surveys and used on simulated data in the framework of survey planning.
\end{abstract}

\firstsection 

\section{The HETDEX survey}

\begin{wrapfigure}{r}{0.45\textwidth}
\includegraphics[width=0.43\textwidth]{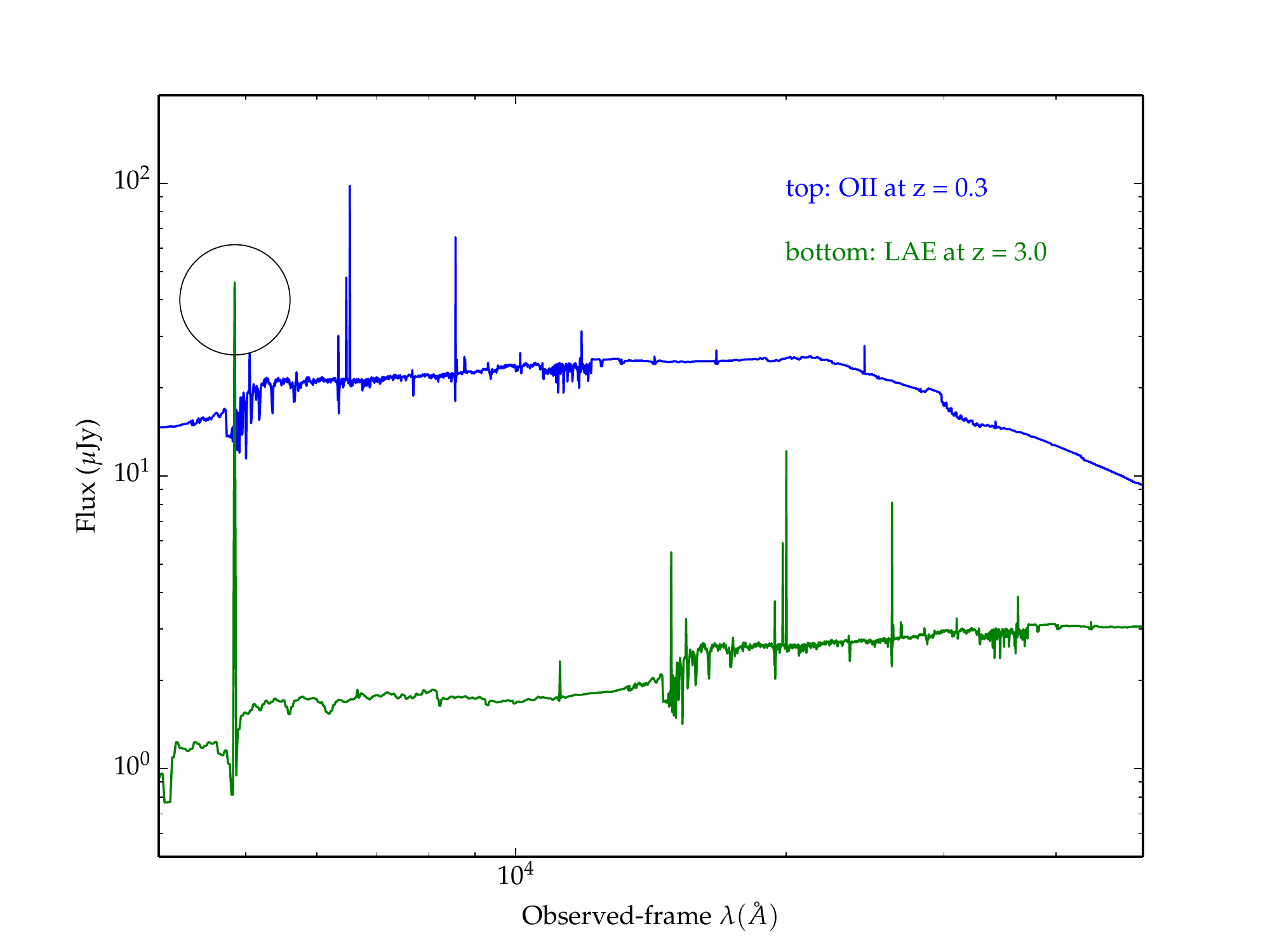}
\caption{\footnotesize{The Lyman-$\alpha$ line from a $z = 3$ galaxy is observed at exactly at the same wavelength as the [OII] line from a $z = 0.305$ galaxy. The emission line flux (shown by the circle) is the same for these two galaxies in this example, but the intensity of the continuum (and therefore the equivalent width, EW) is different. On average, LAEs have much higher EWs than [OII]s.}}
\label{fig:HETDEX}
\end{wrapfigure}

The spectroscopic galaxy survey HETDEX (Hobby-Eberly Telescope Dark Energy eXperiment, http://hetdex.org/) is going to investigate the largely unexplored redshift range of $1.9 < z < 3.5$. By accurately mapping the 3D positions of galaxies, HETDEX will pinpoint the expansion rate of the Universe with a precision of 1\% and detect dark energy, if it exists, at this early epoch. To reach the exciting objective of measuring the expansion rate of the Universe far back in the past, HETDEX will find and catalog several hundred thousand Lyman Alpha Emitting galaxies (LAEs) in the aforementioned redshift range. LAE galaxies are identified through the Lyman$-\alpha$ emission line, a transition between the $n = 2$ and $n = 1$ energy levels of hydrogen in which a photon of wavelength $\lambda = 1216 \AA$ is emitted. Since hydrogen is by far the most common element in the Universe, this high-luminosity line acts as a lamppost even for galaxies very far away. The main source of contamination in HETDEX is an emission line from singly ionized oxygen, [OII], whose rest-frame wavelength is 3727 $\AA$. In the detection range of the HET spectrographs (3500-5500$\AA$), [OII] emitters at redshift $z < 0.56$ might be mistaken for Ly-$\alpha$ emitters, as illustrated in Fig. \ref{fig:HETDEX}. Therefore, a crucial issue involved in maximizing the science impact of the survey is {\bf to make sure that we classify high-redshift LAEs and low-redshift [OII] emitters as accurately as possible.} Towards this end, HETDEX plans a broadband imaging survey in addition to IFU spectroscopy.

\newpage

\section{The name of the game: contamination and incompleteness}

If we were only armed with the measurement of an emission line flux, we would have a hard time distinguishing between Lyman-$\alpha$ and [OII] emitters. Fortunately, there are additional differences between the low-redshift and high-redshift galaxies. The first and most intuitive one is the fact that [OII] emitters, being much closer to us, are on average much brighter. This is usually quantified by the ${\it continuum}$, the density of detected photons with wavelengths near the wavelength of the emission line. The key ratio of (emission line flux)/(continuum flux density)  is called ``equivalent width" (EW):
\be
{\rm EW} = \frac{f_{\rm EL}}{f_{\lambda, {\rm cont}}}
\ee
LAEs have high emission line fluxes and dim continua (because they are farther away), so they have on average much higher equivalent widths. The traditional requirement that LAE have rest-frame EW $>$ 20 $\AA$ (e.g., Gronwall et al. 2007, Adams et al. 2011), 
is effective in limiting the misclassification of [OII] emitting galaxies as LAEs, at the expense of having a highly incomplete sample. Simulations show that this criterion induces a contamination fraction of over $4\%$ at low-redshift.

To describe our challenge more quantitatively, we can define our scientific objective as to minimize the error $\sigma(D_A)$ with which the angular diameter distance to the median redshift of the survey can be measured. This uncertainty depends on two factors: the number of [OII] emitters that are erroneously classified as LAE (contamination), and the number of LAEs that are ``lost" because they are classified as [OII] emitters (incompleteness). Through analysis of errors in the power spectrum reconstruction (E. Komatsu, private comm.) we found that the error in the angular diameter distance can be written as
\be
\sigma_{D_A}^2 \propto \left(\frac{f_{\rm cont}}{0.025}\right)^2 + \frac{270,000}{N_{\rm LAE}}
\label{eq:sigma}
\ee
where $f_{\rm cont}$ is the contamination fraction, and $N_{\rm LAE}$ is the number of recovered LAEs in the survey. This is the quantity that we shall try to minimize in our quest.

\section{Bayesian method}

We developed a technique to separate low- and high- redshift galaxies based on a more comprehensive analysis of the information available in the literature about LAE and [OII] galaxies. The main idea is the following:
\begin{itemize}
\item Start from the luminosity functions and equivalent width distributions for LAE and [OII] galaxies published in Ciardullo et al. (2012) and Ciardullo et al. (2013); 
\item Interpolate those above to obtain a fitting formula valid in the range of redshift of interest;
\item Use these as priors to write a probability P(LAE) (P([OII])) that a galaxy of given observed line flux and continuum flux is an LAE ([OII] emitter), modulo a normalization factor;
\item Consider the ratio P(LAE)/P([OII]) as the main indicator of the classification;
\item Optimize the method by tuning the above ratio as a function of redshift in order to achieve the smallest possible error from Eq. (\ref{eq:sigma}).
\end{itemize}

This technique has been shown to reduce both contamination and incompleteness with respect to the ``$EW > 20 \AA$" criterion. The details will be discussed in an upcoming publication (Leung et al 2014, in preparation).

\section{Machine Learning}

Another approach that we propose here is to classify LAE/[OII] galaxies using supervised {\it machine learning}. Machine learning is the concept of having a computer program ÒlearnÓ about the properties of a dataset by looking for correlations within the different objects in the dataset (called instances).  We plan to use ancillary data existing on some areas of the sky overlapping with the HETDEX survey to identify a subset of galaxies for which a label (LAE or [OII]) can be assigned with a high degree of confidence. This subset is divided into a ``training sample", which is then used to ``learn" what is the best rule to classify an object as an LAE or an [OII] emitter, and a ``validation" sample, which is used to evaluate the performance of the classification procedure and make adjustments as needed.  Contrary to the previous methods, no knowledge of the detailed physical model of galaxies is needed; the algorithm learns the classification rule exclusively by comparison with previous cases for which the correct answer was known. 
In principle, this method can perform better than any of those described above because there is no ``assumption" involved in estimating whether an object is an LAE or an [OII] emitter, and there is no formal limit to how complicated the classification rule can be. On the other hand, the ability of the algorithm to learn the rule correctly depends crucially on the quality and size of the training sample, which needs to be a fair representation of the overall sample of galaxies in order for the method to succeed. \\[-0.2cm]

\underline{\it {Support Vector Machines:}} We chose to use the Support Vector Machine (SVM) classification algorithm (\eg Boser et al. 1992) to decide whether a galaxy is a Lyman Alpha or [OII] emitter. We assume that labels are available for a fraction (up to 1/10th) of the entire dataset.  Each galaxy is an instance, and each instance consists of three attributes, which summarize our anticipated knowledge of the galaxies from the observations, and a label (LAE/[OII]).  The attributes are the observed wavelength of the emission line, the emission line flux, and the continuum flux density.  \\[-0.2cm]

\begin{figure}
\includegraphics[width=0.47\textwidth]{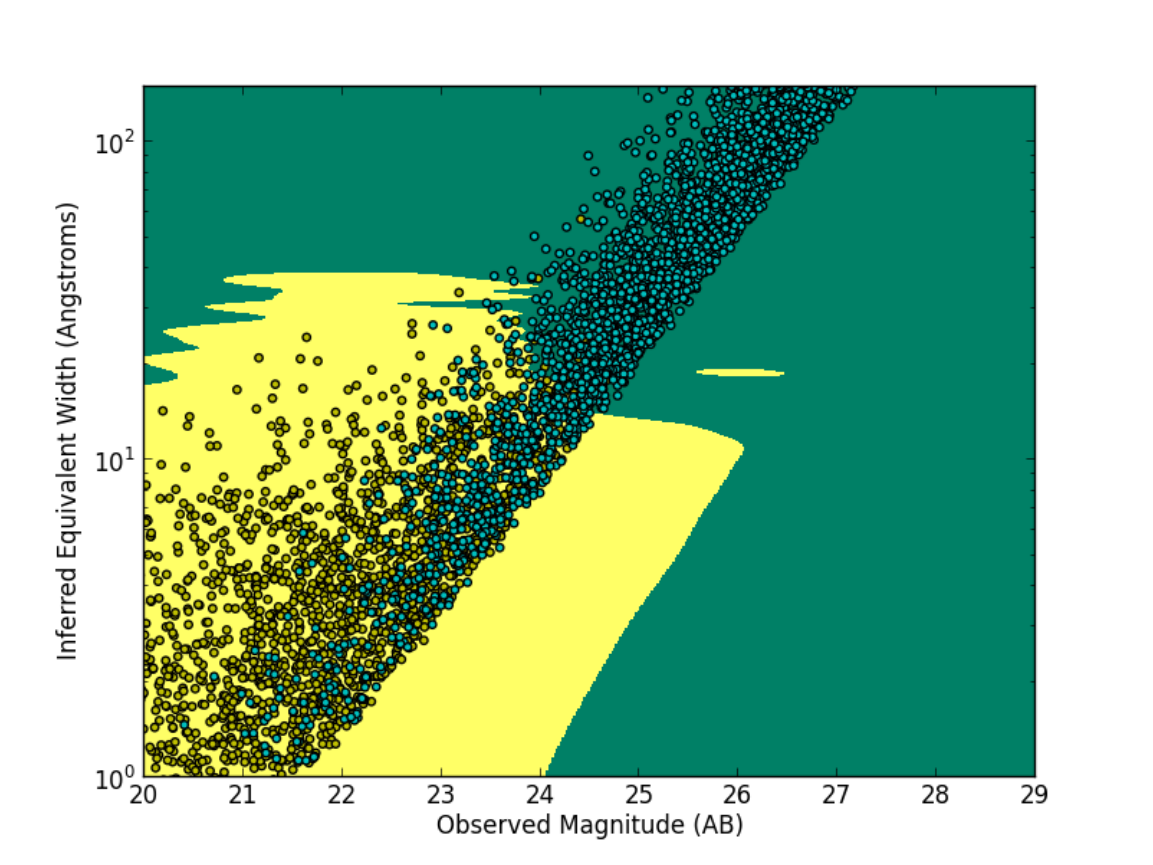}
\includegraphics[width=0.50\textwidth]{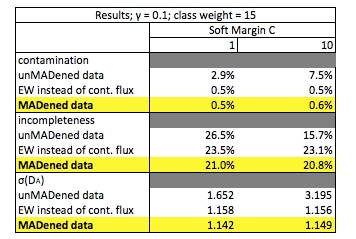}
\caption{\footnotesize{{\it Left:} A graphical representation of our data. Cyan dots represent LAEs and yellow ones represent [OII] emitters.  The separation between colors is the decision boundary. Any objects that falls into the blue area is classified by the SVM as an LAE. Note: this is a simplified version with a 2D decision boundary for illustration purposes. {\it Right:} A summary of the results obtained by optimizing the SVM algorithm. The numbers for $\sigma(D_A)$ are {\it proportional to} the actual measurement error. }}
\label{Fig:SVM}
\end{figure}

\underline{\it {How SVM Works:}} SVM works by finding a hyperplane within the dataset that separates the two classes (see Fig. \ref{Fig:SVM}).  In order to find this hyperplane in a non-linearly-separable dataset, the algorithm must transform the data into higher dimensions.  This is done through different mappings, often called kernels.  The kernel that we have found to be most successful is the radial basis function (rbf) kernel.  This kernel is Gaussian, and has several parameters which can be adjusted: \\[-0.2cm]
\begin{enumerate}
\item The amplitude parameter $\gamma$, which defines the complexity of the decision boundaries.  
\item The soft margin constant C. 
Margins are defined by maximizing the distance between the boundary hyperplane and the closest objects in the dataset (also called support vectors). 
A soft margin, as is used in the rbf kernel, allows for some error.  For example, if one instance is beyond the margin, one can allow for that instance to be misclassified if this overall increases the performance of the algorithm. A large value of C corresponds to a small tolerance for these misclassification. 
\item
The class weight. Adjusting this parameter corresponds to assigning a higher penalty to misclassifications of one type of object (in practice, a different value of C according to the class).\\[-0.2cm]
\end{enumerate}

\underline{\it {Results:}} Through a grid search of all possible values of C, $\gamma$, and class weight, we found the most efficient hyperplane, as shown in the Table of Fig. \ref{Fig:SVM}.  The results for soft margin values of C = 1 and C = 10 are both shown since they are close to optimal. Our best values of $\gamma$ and class weight were 0.1 and 15 respectively. Other steps in the optimization process consisted of ``standardizing" data (or more precisely, ``MADening" the data, by rescaling them to all have the same median and median-based-deviation), and checking which combination of attributes delivers the best results.

\section{Work in progress}

Both the Bayesian method and SVMs showed substantial improvement with respect to the EW = $20 \AA$ cut. We plan to keep optimizing both techniques by:
1. including information about the color in the SEDs; 2. extending the simulations to include other emission lines that can help the classification of [OII] emitters (such as the other emission lines in the Balmer series, and [NeIII]); 3. considering other machine learning algorithms, such as boosted decision trees and random forests, in addition to SVMs.

\end{document}